# Low Background Stainless Steel for the Pressure Vessel in the PandaX-II Dark Matter Experiment

**Tao Zhang,** [a] **Changbo Fu,** [a] **Xiangdong Ji,** [a, b, 1] **Jianglai Liu,** [a] **Xiang Liu,** [a] **Xuming Wang,** [a] **Chunfa Yao,** [c] **Xunhua Yuan** [c]

[a] *INPAC and Department of Physics and Astronomy, Shanghai Jiao Tong University, Shanghai Laboratory for Particle Physics and Cosmology, Shanghai 200240, China*
[b] *Department of Physics, University of Maryland, College Park, Maryland 20742, USA*
[c] *China Iron & Steel Research Institute Group, Beijing 100081, China*
*Email: tzhang@sjtu.edu.cn, xdji@sjtu.edu.cn*

ABSTRACT: We report on the custom produced low radiation background stainless steel and the welding rod for the PandaX experiment, one of the deep underground experiments to search for dark matter and neutrinoless double beta decay using xenon. The anthropogenic $^{60}$Co concentration in these samples is at the range of 1 mBq/kg or lower. We also discuss the radioactivity of nuclear-grade stainless steel from TISCO which has a similar background rate. The PandaX-II pressure vessel was thus fabricated using the stainless steel from CISRI and TISCO. Based on the analysis of the radioactivity data, we also made discussions on potential candidate for low background metal materials for future pressure vessel development.

---

[1] Corresponding author

**Contents**

## 1. Introduction

PandaX is a large upgradable xenon detector system to search for dark matter and neutrinoless double beta decay operating in China Jin-Ping Underground Laboratory (CJPL) [1]. The experiment has completed the first-phase mission (PandaX-I) with 120 kg of liquid xenon target mass [2] [3], and the second phase, PandaX-II, with more than a factor of four increase in target mass, is under commissioning in CJPL. To achieve lower background level, such experiment has extremely stringent requirements on the radio purity of materials [4]. In particular, liquid xenon must be stored in a pressure vessel of excellent radio purity due to its proximity to the target region.

In general, oxygen-free high thermal conductivity copper (OFHC) has been widely used in low background experiment [4] [5] because of its high radio purity. However, the mechanical properties of OFHC are poor and it is difficult to weld. On the other hand, stainless steel (SS) is a common material in pressure vessel and vacuum industry. High-level radiopurity for SS is also feasible, e.g. the GERDA experiment has identified such samples [6].

In this paper, we report the production of SS with similar radioactivity as the steel used in GERDA experiment. In Sec. 2, we discuss radiation assays of various samples. Two clean research-grade samples from China Iron & Steel Research Institute Group (CISRI) were found. In Sec. 3, we report the smelting process at CISRI. In Sec. 4, the industrial production of radio-pure nuclear-grade SS from Taiyuan Iron & Steel (Group) Co. LTD (TISCO) is discussed. In Sec. 5, the pressure vessel fabricated for PandaX-II by using clean SS is presented in details. Based on the measurements of our samples, we also discuss the origin of the anthropogenic radioactivities and propose potential candidates for future low background pressure vessels in Sec. 6.

## 2. SS requirements and sample assay

Monte Carlo simulations were carried out to control the radiation contributions from the pressure vessel under 0.1 mDRU (1 DRU = 1 event/kg/day/keV) in the fiducial volume of the PandaX-II detector. Assuming the radioactivity level of each isotope is 1 mBq/kg, the background budgets from different parts of the pressure vessel are listed in Table 1. It is then easy to estimate whether a sample satisfies our requirements.

Table 1: Preliminary vessel design and resulting background (0.001 mDRU) in the fiducial volume of the detector assuming that the radioactivity of the isotopes in all materials is 1 mBq/kg.

| Isotope(asumse 1 mBq/kg ) | $^{238}$U | $^{232}$Th | $^{235}$U | $^{137}$Cs | $^{60}$Co | $^{40}$K |
|---|---|---|---|---|---|---|
| Barrel and endcaps ~190 kg | 6.2±1.2 | 2.8±0.8 | <0.49 | 1.9±0.6 | 5.2±1.1 | 0.21±0.21 |
| Flanges ~100 kg | 2.1±0.5 | 0.8±0.3 | <0.25 | 0.11±0.11 | 2.5±0.5 | 0.22±0.15 |
| Overflow chamber ~20 kg | 0.21±0.07 | 0.24±0.07 | 0.021±0.021 | <0.049 | 0.043±0.030 | 0.043±0.030 |

The PandaX collaboration has setup a gamma counting station at CJPL with a sensitivity of 1 mBq/kg level [1], the results of these samples are summarized in Table 2. For all random samples found in the market, although the radioactivity level varied, the overall cleanliness did not satisfy our requirements based on the simulation results of Table 1. This is similar to the results reported in Ref. [6] [7]. The only two exceptions are the samples (S-I and S-J) from the CISRI [8], which is a professional institute that carries out metal-related research. S-I is SS316L and S-J is nichrome.

Table 2: The radiopurity assay results of different samples. The S-J sample is nichrome and others are SS. Radionuclide concentrations in the samples are given in unit mBq/kg.

| Sample | $^{226}$Ra ($^{238}$U) | $^{228}$Ac ($^{232}$Th) | $^{228}$Th ($^{232}$Th) | $^{235}$U | $^{210}$Pb | $^{137}$Cs | $^{60}$Co | $^{40}$K |
|---|---|---|---|---|---|---|---|---|
| S-A | 4.8±1.0 | 9.2±2.1 | 12±2 | 6.7±2.3 | <306 | <0.45 | 13±1 | 17±9 |
| S-B | 33±2 | <1.9 | 6.3±1.7 | 6.7±4.6 | <804 | <0.61 | 6.5±0.8 | <7.1 |
| S-C | 56±2 | 10±2 | 13±2 | <3.4 | <510 | 5.6±0.7 | 13±1 | 12±7 |
| S-D | <1.7 | 4.2±2.7 | <2.2 | 4.9±2.7 | <504 | 1.0±0.8 | 9.0±1.0 | <13 |
| S-E | <3.0 | 12±3 | 14±3 | <3.7 | 562±220 | 79±2 | 7.0±1.0 | 31±14 |
| S-F | 36±2 | 2.6±1.6 | 6.8±1.6 | <2.4 | 526±225 | <0.42 | 19±1 | <6.7 |
| S-G | 62±3 | 6.3±3.0 | 6.3±2.9 | <5.3 | <578 | 1.1±0.9 | 75±2 | <16 |
| S-H | 118±3 |  | 17±4 | <8 |  | <0.9 | 6.0±0.8 | 53±14 |
| S-I | <1.6 | <2.5 | <2.2 | 3.2±2.5 | <436 | <0.96 | <0.56 | <11 |
| S-J | <1.5 | 2.8±2.3 | 10±3 | <4.8 | <616 | 2.7±0.8 | <0.52 | <11 |

It is common to use scrap materials in the smelting process to produce SS [6]. Therefore it is easy to understand that the radiopurity cannot be well controlled. CISRI, on the other hand, starts from very pure raw materials, which might make the key difference.

## 3. Low background SS smelting process

In order to produce radio pure steel, we controlled the entire smelting process as much as we could together with CISRI. First, since two of the most important ingredients in SS are Fe and Ni, we measured the radioactivity of pure iron (S-1) from TISCO [9] and electrolytic nickel (S-2) from Jinchuan Group Co. LTD [10]. As expected, the radiopurity satisfies our requirements.

Table 3: The radiopurity assay results of different samples. S-1 sample is pure iron, S-2 sample is electrolytic nickel, and others are SS. Radionuclide concentrations in the samples are given in unit mBq/kg.

| Sample No | $^{226}$Ra ($^{238}$U) | $^{228}$Ac ($^{232}$Th) | $^{228}$Th ($^{232}$Th) | $^{235}$U | $^{210}$Pb | $^{137}$Cs | $^{60}$Co | $^{40}$K |
|---|---|---|---|---|---|---|---|---|
| S-1 | <0.5 | <1.2 | <0.6 | <0.2 | | <1.6 | <0.75 | <2.2 |
| S-2 | 1.1±1.1 | <1.9 | <1.2 | <0.47 | | 5.5±3.0 | 1.7±1.3 | <4.4 |
| S-3 | <1.3 | <1.9 | <1.5 | <2.4 | 382±294 | 0.60±0.58 | <0.53 | <13 |
| S-4 | <1.5 | <1.6 | <1.3 | <1.7 | <274 | 0.49±0.45 | <0.40 | 9.2±8.8 |
| S-5 | 5.3±2.0 | 11±3 | 5.1±2.6 | <2.6 | <336 | 2.9±0.9 | <0.74 | <15 |
| S-6 | <2.4 | <4.1 | <2.2 | 3.8±3.0 | <300 | <1.1 | <0.94 | <19 |
| S-7 | 5.6±4.2 | <5.6 | <2.5 | <1.9 | <520 | <1.8 | <1.5 | <35 |
| S-8 | 8.8±1.7 | 39±4 | 31±3 | <2.4 | <310 | 1.5±0.7 | 1.7±0.7 | <11 |
| S-9 | <1.4 | <2.4 | <1.9 | <2.0 | <196 | <0.68 | 0.75±0.63 | <12 |
| S-10 | 1.3±1.1 | <1.7 | 6.1±1.9 | <1.7 | <239 | <0.47 | 0.97±0.45 | <8.1 |
| S-11 | <1.7 | <2.7 | <1.7 | <2.4 | <241 | 2.4±1.0 | 1.0±0.8 | <14 |
| S-12 | <1.9 | <3.0 | <3.4 | <2.7 | <326 | 1.4±1.0 | <0.7 | <16 |
| S-13 | 15±11 | <18 | <9.4 | <12 | <1440 | <4.4 | <3.7 | 78±77 |

We then smelted 18 kg SS304L at CISRI as a pilot sample. The raw materials used and their purity are listed in Table 4. Compared with SS316L, the chemical composition of SS304L is simpler, so SS304L may be less prone to radio-contamination. The pure iron was from TISCO, and the others were electrolytic refining materials. The impurities in raw material can be illustrated by the electrolytic nickel, the purity 0.9996 means the mass percentage of Ni and Co should be more than 99.96% and Co mass percentage should be less than 0.02%, and other elements, e.g. C, Fe, Si, Mn, Al, Cu, Cr, As, Cd, Sn, Sb, Zn, Mn, Mg, Bi, P, S etc, should be less than some certain value [12]. The other elements e.g. U, Th, Ra are too low to be indicated in metallurgical industry. Of course, some "impurities", like Fe, Cr, Mn, are not considered as "impurities" for SS304L. And so are other impurities in other raw materials.

The smelting was performed in a MgO crucible in a vacuum chamber with an ingot mould next to it, which is shown in Figure 1. To ensure the cleanliness, we also checked the crucible log and made sure that no cobalt had been melted in it. We summarize the main steps of the metallurgy process as follows:

1) Flushing the crucible: load pure iron piece into the crucible and pump the chamber to 10Pa. Then melt the iron in the crucible and remove the liquid iron afterwards to "clean" the surface;

2) Dedust the pure iron, Cr, and Ni, and load them into the crucible. Pump the chamber to less than 10 Pa and then heat and melt the raw materials;

3) Continue pumping to 1 Pa to degas the $O_2$, $N_2$, and $H_2$;

4) In the vacuum chamber, load the rest raw material into the crucible and mix the liquid metal;

5) Cast the alloy into ingot mould;

6) Vent and open vacuum chamber and take the steel ingot out.

Table 4: 18kg SS304L ingot raw materials list

| Raw material | Mass(g) | Purity |
|---|---|---|
| Fe (Pure Iron) | 12616 | 0.997 |
| Cr | 3449 | 0.993 |
| Ni | 1620 | 0.9996 |
| Si | 90 | 0.996 |
| Mn | 180 | 0.999 |
| Al | 18 | 0.999 |
| Ni-Mg | 27 | |

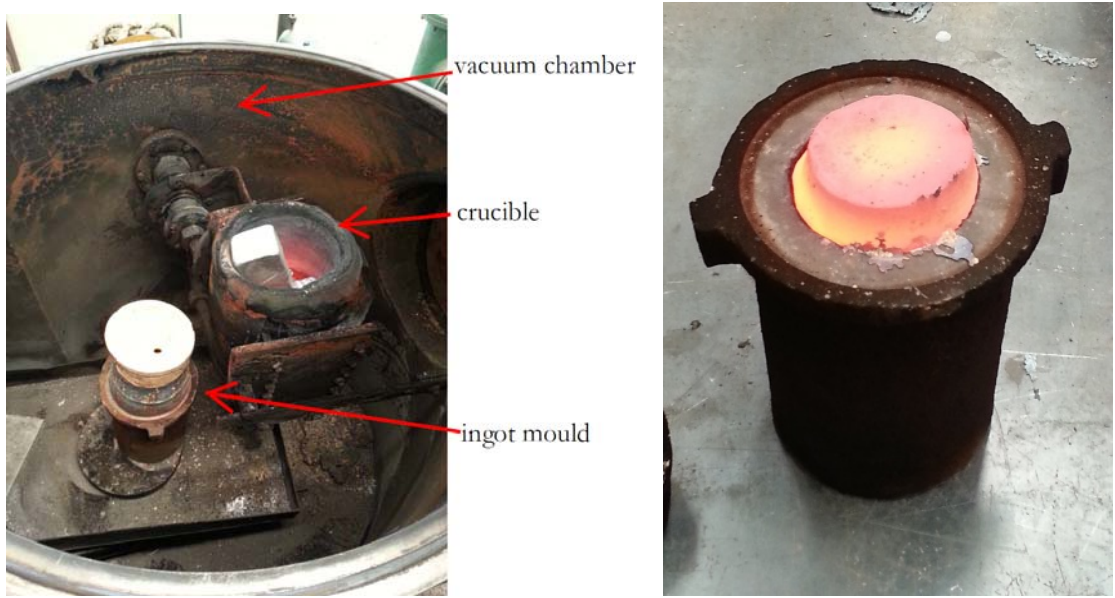

Figure 1: Vacuum furnace, crucible and SS304L ingot

In Table 3, S-3 and S-4 are samples from the top and bottom of the ingot. The smelting process did not introduce additional radioactivities. Similar to Ref. [6], we do not observe obvious inhomogeneous distribution of radio nuclides in the ingot. The alloy's chemical compositions were measured with a metallurgical spectrometer, as is shown in Table 7. Besides the elements in Table 7, Ti(0.017%), V(0.012%), Nb(0.0067%), Mo(0.0041%), Sn(0.0051%) and W(0.036%) were found too. The elements of U, Th, Ac, Cs, Ra and K were not detected, i.e. their concentrations were below the sensitivity limits of the metallurgical spectrometer.

Given that the small pilot sample worked successfully, 3-ton of SS304L were made with the same raw materials and the same smelting procedure in a larger and brand new crucible with 1000-kg capacity. Nine 300-kg ingots were made. S-5 and S-6 samples in Table 3 were taken from different ingots. It was found that the radioactivity level of S-5 and S-6 was, for reasons beyond our understanding, higher than that of S-3 and S-4. The ingots were forged into different shapes including 40 mm thick 930 mm outer diameter flanges, plates with different thickness, rods, bolts and nuts. The S-7 in Table 3 is a sample from the plates, indicating that the forging process did not introduce significant radioactivities.

To ensure the radio-purity of the welds, standard welding rods AT-ER316L (2.5mm diameter, 500 g) from CISRI were also checked (S-8 in Table 3). The $^{232}$Th concentration was high, but the weld joint mass was about 1% of the vessel, which was acceptable. The raw material of this type of welding rod and the smelting process were very similar to samples S-3 and S-4. The same batch of welding rods was used in the vessel fabrication.

## 4. Low background commercial nuclear grade SS

The nuclear-grade SS is a special type of SS used in nuclear reactors. To avoid neutron activation, the mass percentage of Co should be less than 0.05%. To produce the nuclear-grade SS, liquid cast iron and electrolytic refined Ni, Cr, Mn, Si and Al are used. The liquid cast iron is produced in a blast furnace directly from the iron ore, coal, etc., and avoiding the use of recycled scrap material. Both steps help to obtain low radioactivity final products. Compared to CISRI, the large-scale production process in TISCO has the following key differences (see schematic diagram in Figure 2). First, liquid cast iron is used instead of the pure iron. Pure iron is produced from the cast iron by blowing pure oxygen in crucible at high temperature, and then has less impurity. Second, the crucible of TISCO has a 40~50 ton capacity, which has been used many times in the past. Normally there is also no flushing step to clean the crucible. Third, the alloy materials such as Ni and Cr are stored outdoors without special precautions. They are loaded into the crucible without dedusting. However, low radioactivity SS could be produced, two samples produced using such process with 3 mm and 6 mm thickness respectively (S-9 and S-10 in Table 3) were measured to be very low in radioactivity.

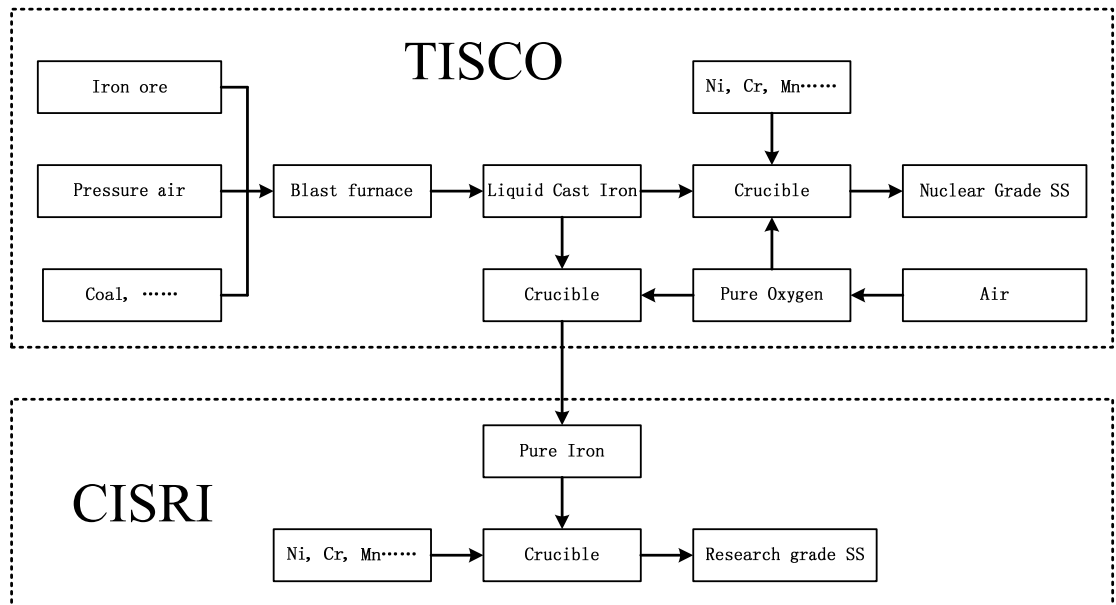

Figure 2: Stainless steel making flow chart

Given the measurement results, TISCO donated two nuclear grade SS304L plates, each from a different production batch, to the PandaX collaboration, with dimensions of 6000×1300×3 mm and 4000×2000×6 mm, respectively. S-11 and S-12 in Table 3 are corresponding samples from these two plates – their radioactivities were satisfactory as well.

## 5. PandaX-II Pressure Vessel fabrication

The PandaX-II liquid xenon cryostat is placed in an outer vacuum vessel made of pure Cu. The typical operating pressure for the PandaX-II vessel is 2 bar. Therefore, the inner vessel has to be a high vacuum vessel, a pressure vessel rated 0.35 MPa , and a cryogenic vessel with proper seals functioning at -100℃.

A design drawing and photo of this vessel are shown in Figure 3. The inner diameter is 800 mm. The barrel height is about 1085 mm with a dome welded in the bottom. The top lid is also a dome with a 40 mm thick flange. The seal between the top flange and barrel is an indium wire of 3 mm diameter. The permissible stress of SS304L is 120 MPa according to Chinese national standard GB 150-2011 pressure vessels [11]. To hold vacuum, a 6 mm thickness is required for the barrel and endcaps, although only 2 mm are sufficient for the upper inner pressure of 0.35 MPa. The overall vessel weighs about 310 kg, and a summary of part weights is listed in Table 7. An overflow chamber, to control the liquid xenon level, is located in the inner vessel, about 5 mm away from the bottom endcap. It weighs 18.6 kg, and volume is 18 L.

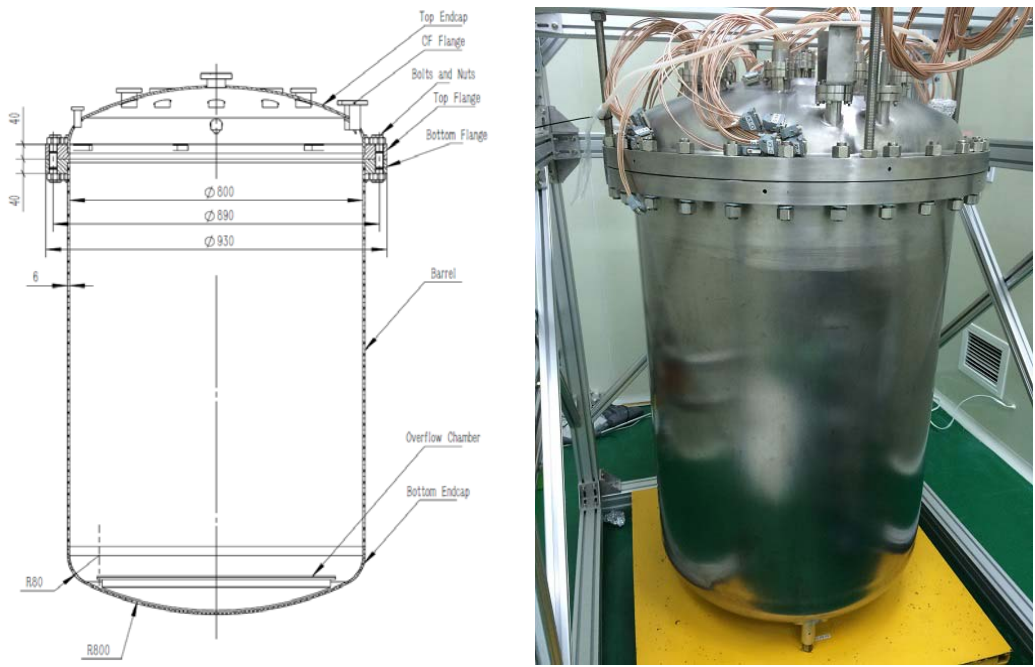

Figure 3: PandaX-II Pressure Vessel Drawing and photo

Using low background SS and welding rod (S-12 and S-8 in Table 3), we made an argon-arc welding test with a WC20 electrode (alloy of W and 1.8%~2.2% Ce). The sample (S-13) has dimensions of $100\times10\times7$ mm$^3$ and weight of 54 g, about 20 g of which are from the welding rod. Due to the low weight of the sample, the counting results had large uncertainties and were mostly upper limits. Consequently, no radio-impurity was identified after this process.

The PandaX-II pressure vessel was fabricated using these materials. The corresponding samples from the flanges, endcaps, barrel, as well as bolts and nuts are summarized in Table 5. The average welded joint contains about 0.2 kg/m of welding material, and in total about 3 kg for the entire pressure vessel.

Table 5: Vessel parts weight list

| Item | Weight (kg) | Corresponding Sample |
|---|---|---|
| Top & bottom flanges | 49×2 | S-6 |
| Barrel | 124 | S-12 |
| Endcaps | 33×2 | S-12 |
| Bolts and nuts | 16 | S-7 |
| CF flanges | 6 | S-7 |
| Pressure Vessel (total) | 310 | |
| Overflow chamber | 18.5kg | S-11 |

## 6. Discussion of the results

Using the data shown in the Table 1, Table 3 and Table 5, the radioactive background from the pressure vessel was calculated to be less than $0.052\pm0.006$ mDRU. The details are shown in Table 6. In 2015 and 2016, a series of runs were carried out, and a WIMP search data set with an exposure of $3.3\times10^4$ kg·day was taken. The background level in Table 6 was used as an input component of the total expected background[14] [15].

Table 6: PandaX-II pressure vessel and overflow chamber background (mDRU). The radionuclide concentrations of the vessel parts are given in unit mBq/kg.

| Component | Mass(kg) | $^{238}$U | $^{232}$Th | $^{235}$U | $^{137}$Cs | $^{60}$Co | $^{40}$K | Background |
|---|---|---|---|---|---|---|---|---|
| Vessel barrel and endcaps | 190 | <1.9 | <3.0 | <2.7 | 1.4±1.0 | <0.7 | <16 | <0.030±0.005 |
| Top & bottom flange | 98 | <2.4 | <4.1 | 3.8±3.0 | <1.1 | <0.9 | <19 | <0.015±0.003 |
| Overflow Chamber | 18.5 | <1.7 | <2.7 | <2.4 | 2.4±1.0 | 1.0±0.8 | <14 | <0.0016±0.0005 |
| Bolts and nuts | 8 | 5.6±4.2 | <5.6 | <1.9 | <1.8 | <1.5 | <35 | <0.0015±0.0006 |
| Bolts and nuts ‡ | 7 | <15 | <6 | <9.4 | <6 | <2.8 | <99 | <0.0036±0.0009 |
| Total | | | | | | | | <0.052±0.006 |

‡ Due to strength limitation of SS304L bolts, half of bolts and nuts are replaced by comerical ones in the commision.

Compared with the samples of the GERDA experiment [6], XENON100 experiment [5], and LZ experiment [16], our SS has lower concentrations of $^{60}$Co (~1 mBq/kg) since our samples started with cast iron without scrap materials. For all the other isotopes, our upper limits are not as constrained as the ones of GERDA due to the sensitivity of our counting station.

To see if it is possible to reduce the radioactivity of SS further, we discuss the potential for producing purer iron. Iron can be purified by electrolytic refining to 0.999 to 0.9999 purity, but costing more than Cu and Ni with the same purity [12]. Considering that Cu is less radioactive and nickel alloy has better mechanical property and corrosion resistance, electrolytic iron is probably not cost effective. Another potential improvement would be to refine the iron-making process, e.g. the high radioactive coal [17] can be replaced by $H_2$ or CO as deoxidizer, which could lead to lower introduction of radionuclides.

The pure iron has another potential application. In normal low background experiments, Pb bricks are used as gamma shield [1]. On the other hand, pure iron has better radiopurity and much better mechanical properties, despite the lower atomic mass. The clean materials in Table 3 contain C, Si, Fe, Mn, Al, Cr, Ni elements, indicating that they would not introduce much radio-impurity. So besides SS, other low background alloys using these elements are also quite possible. For example, carbon steel consists of Fe, C, Mn and Si, which has fewer alloy elements than SS. Futhermore, TISCO can produce nuclear-grade carbon steel with a relatively low price, which could be used massively in low background experiments as shielding material.

It is interesting to analyze the chemical composition of some clean samples that we have studied earlier (see Table 3). In China, most nickel and cobalt are commonly associated [13]. Interestingly, in samples S-J and S-2 with very high Ni concentration, the $^{60}$Co background does not seem to be different from other samples, as shown in Table 7. This implies that nickel might not be the only source of $^{60}$Co contamination. For applications requiring even lower background, the Co concentration in 0.9999 purity electrolytic nickel can be less than 50 ppm [14]. Besides the electrolytic refining process, higher purity nickel can be produced by carbonyl process refinement, which helps to remove radio-impurities. For example, the Co could be reduced to 3 ppm in Ref. [18], and a carbonyl nickel pellet sample made with the same technique is reported to have 0.4 mBq/kg, 0.09 mBq/kg and 0.4 mBq/kg for $^{40}$K, $^{232}$Th and $^{238}$U respectively [19].

Table 7: Sample chemical composition (%)

| Sample | Fe | Cr | Ni | Si | Al | C | Mn | P | S | Cu | Co | B | N | Pb |
|---|---|---|---|---|---|---|---|---|---|---|---|---|---|---|
| S-I | 65.13 | 17.01 | 12.73 | 0.48 | 0.13 | | 0.95 | 1.65E-2 | 5.50E-3 | | 2.30E-3 | | | |
| S-J | 0.05 | 19.67 | 79.58 | 0.20 | 0.20 | | 0.00 | 2.00E-3 | 6.60E-3 | | 1.26E-2 | | | |
| S-3/S-4 | 70.20 | 18.63 | 9.29 | 0.57 | 0.10 | 3.77E-3 | 1.00 | 7.25E-3 | 8.94E-3 | 9.30E-2 | 1.44E-2 | | | |
| S-5 | 70.53 | 18.78 | 8.99 | 0.53 | 0.09 | 1.11E-2 | 0.99 | 5.82E-3 | 3.73E-3 | 1.20E-2 | 4.90E-3 | 2.70E-4 | | 8.64E-6 |
| S-6 | 70.50 | 18.94 | 8.91 | 0.53 | 0.09 | 1.42E-2 | 0.99 | 5.90E-3 | 3.69E-3 | 1.30E-2 | 4.86E-3 | 2.20E-4 | | 6.69E-6 |
| S-7 | 70.64 | 18.97 | 8.90 | 0.51 | 0.02 | 7.95E-3 | 0.91 | 5.95E-3 | 3.58E-3 | 1.26E-2 | 4.80E-3 | 2.60E-4 | | 1.97E-5 |
| S-8 | | 18.63 | 12.25 | 0.57 | | 9.00E-3 | 2.19 | 1.60E-2 | 7.00E-3 | 2.00E-2 | | | | |
| S-11 | | 18.30 | 9.03 | 0.56 | | 2.20E-2 | 1.07 | 2.50E-2 | 1.00E-3 | 2.98E-2 | 3.06E-2 | 3.00E-4 | 5.00E-2 | |
| S-12 | | 18.20 | 9.11 | 0.53 | | 2.00E-2 | 1.14 | 2.30E-2 | 1.00E-3 | 2.55E-2 | 3.03E-2 | 3.00E-4 | 5.00E-2 | |

OFHC usually has the lowest radioactive contamination in metal materials avaiable in ordinary markets [6] [20], but its mechanical properties are poor (allowable stress is 27MPa in copper pressure vessels design code [22]). Furthermore copper is difficult to be welded. In general, the mechanical properties of alloys are better than those of pure metals and radio-pure bronze (Sn-Cu alloy) has been successfully used in the CRESST experiment [21]. Brass and white copper are feasible too for pressure vessels [22]. For example, H96 brass consists of 95%~97% copper and 3~5% Zn, and its allowable stress in pressure vessel is 40 MPa [22], about 1.5 times that of OFHC's. Zinc can be purified either by electrolytic refining or rectification to a higher purity than the copper cathode [23] [24], as well as by distillation [18] [23]. The chemical activity of zinc is in between those of Cu and Al, therefore we suspect that the Th, U and K concentrations in zinc may be in between the electrolytic Al [25] and copper cathode. The decay products of

$^{64}$Zn and $^{70}$Zn are either electron or positron, which are relatively easy to be shielded. White copper is an alloy of copper and nickel with much better mechanical properties than copper. For example, B19 (18%~20% Ni and 80%~82% Cu) has allowable stress of 80 MPa [22]. NCu30 (ISO NW4400, ASTM N04400), contains 28%~34% Cu and >63% Ni (other elements are unnecessary) [26]. Its allowable stress is 172 MPa in pressure vessel [27], which is about 6.3 times that of OFHC. Since high radiopurity nickel is feasible based on our measurements, these alloys are potential candidates for structural materials in ultra-low background experiments.

Besides alloys, clad plate is other candidate for low background design, and it is commonly used in pressure vessels. In general, the SS layer thickness is about 3 mm in clad plate for an ordinary pressure vessel. According to the PandaX-II vessel design requirements, a clad plate with 3 mm thickness SS inner layer (for sealing, corrosion resistance and smoothness requirements) and 4 mm thickness OFHC outer layer is sufficiently strong and stiff. According to the Monte Carlo simulation shown in Table 6, if the barrel and endcaps are made of SS-Cu clad plate, the pressure vessel background could be reduced to 0.037 ±0.004mDRU from 0.052 ±0.006mDRU.

## 7. Summary

In this paper, we report a process to produce low background stainless steel in a small-scale research setup, as well as in a large-scale commercial steel making setup. Our measurement shows that SS samples produced from both setups were not contaminated by anthropogenic radionuclides, especially $^{60}$Co, which were significantly lower than in most previous reports. A low background pressure vessel was fabricated for the PandaX-II dark matter experiment using low background SS. Based on our results, it is very likely that ultra-low radioactive background materials for rare event search experiments could be produced with these processes.

## 8. Acknowledgments

This project has been supported by a 985-III grant from Shanghai Jiao Tong University, grants from National Science Foundation of China (Grants No. 11435008, No. 11455001, and No. 11525522), a grant from the Ministry of Science and Technology of China (Grant No. 2016YFA0400301), and a grant from the Office of Science and Technology in Shanghai Municipal Government (Grant No. 11DZ2260700). This work is supported in part by the Chinese Academy of Sciences Center for Excellence in Particle Physics (CCEPP). We are particularly indebted to Director De Yin from TISCO for crucial help on nuclear-grade steel plates. We also thank Shanghai KingSun Steel Product Co.LTD for pressure vessel design and fabrication.